\documentclass[a4paper, 12pt]{article}

\usepackage{amsmath,amsthm}

\usepackage{etex}
\usepackage{bm, bbm}
\usepackage{amssymb}
\usepackage{mathtools}
\usepackage{natbib}
\usepackage{epsfig}
\usepackage{pdflscape}
\usepackage{sectsty}
\usepackage{graphicx}
\usepackage[margin=1in, nohead]{geometry}
\usepackage{hyperref}
\usepackage{setspace}
\usepackage{caption}
\usepackage{dcolumn}
\usepackage{booktabs}
\usepackage{xr}
\usepackage{soul}
\usepackage{calc}
\usepackage{epigraph}
\usepackage{enumitem}
\usepackage{abstract}
\usepackage{pgfplots}
\usepackage{tikz}
\usetikzlibrary{decorations.pathreplacing}
\usetikzlibrary{calc}
\usepackage{footmisc}
\usepackage{titlesec}
\usepackage{array}
\usepackage{placeins}
\newcolumntype{L}[1]{>{\raggedright\let\newline\\\arraybackslash\hspace{0pt}}m{#1}}
\newcolumntype{C}[1]{>{\centering\let\newline\\\arraybackslash\hspace{0pt}}m{#1}}
\newcolumntype{R}[1]{>{\raggedleft\let\newline\\\arraybackslash\hspace{0pt}}m{#1}}
\newcolumntype{d}[1]{D{.}{.}{#1}}

\setlist[itemize]{leftmargin=*}

\titlelabel{\thetitle.\quad}
\titleformat*{\section}{\centering\scshape\large}
\titleformat*{\subsection}{\centering\itshape}
\titleformat{\subsubsection}[runin]{\bfseries}{\thesubsubsection}{1em}{}

\graphicspath{{./}}

\hypersetup{
    colorlinks,%
    citecolor=brown,%
    filecolor=blue,%
    linkcolor=blue,%
    urlcolor=black
}

\graphicspath{{figures/}}

\newcommand{\I}{\mathbbm{1}}

\allsectionsfont{\raggedright\mdseries}\subsectionfont{\raggedright\itshape\mdseries}
 
\usetikzlibrary{fit,positioning}
\usetikzlibrary{decorations.pathreplacing}
\usetikzlibrary{calc}

\begin{document}

\title{{\bf Modeling Asymmetric Relationships \\ from Symmetric Networks}}

\author{
  Arturas Rozenas\\
{\small NYU}\\
  {\small \texttt{ar199@nyu.edu}}
\and
  Shahryar Minhas\\
{\small MSU}\\
  {\small \texttt{minhassh@msu.edu}}
  \and
  John Ahlquist\\
{\small UCSD}\\
  {\small \texttt{jahlquist@ucsd.edu}}
\thanks{Versions of this paper were presented at the 2015 meetings of the International Political Economy Society, the 2016 meetings of the Society for Political Methodology, and WardFest.  We thank James Fowler, Jenn Larson, and Mike Ward for useful comments. Micah Dillard provided excellent research assistance.  Ahlquist benefitted from a fellowship at Stanford's Center for Advanced Study in the Behavioral Sciences during the writing of this paper.  Installation instructions for the P-GBME package and files to replicate the analyses in this paper are available at \url{http://github.com/s7minhas/pgbmeRepl}.}}

\maketitle 

\begin{center}
  \Large{Forthcoming Political Analysis}
\end{center}

\thispagestyle{empty}

\bigskip

\begin{abstract}
 \noindent Many bilateral relationships requiring mutual agreement produce observable networks that are symmetric (undirected). However, the unobserved, asymmetric (directed) network is frequently the object of scientific interest. We propose a method that probabilistically reconstructs the latent, asymmetric network from the observed, symmetric graph in a regression-based framework. We apply this model to the bilateral investment treaty network. Our approach successfully recovers the true data generating process in simulation studies, extracts new, politically relevant information about the network structure inaccessible to alternative approaches, and has superior predictive performance. 
\end{abstract}

\clearpage

\singlespacing

\section{Introduction} 

Social actors are often embedded in webs of relationships that profoundly shape political and economic outcomes \citep{franzese:hays:2008,ward:etal:2011}. One challenge in analyzing networks arises in situations where an analyst cannot fully observe the nature of relational ties. In many dyadic interactions --- treaties, marriages --- outsiders can observe ties only if both agents agree, that is, the payoff for forming a tie exceeds its cost for \emph{both} members of a dyad \citep{jackson:wolinsky:1996}. The observed network is therefore composed of symmetric (``undirected'') ties even though the social process at work contains important relational asymmetries. The pursuit of a tie by one party may not be reciprocated to the same extent by another.

As an illustration, suppose $A$, $B$, and $C$ are three warring factions deciding whether to sign bilateral peace agreements. We observe a network in which dyads $(A, C)$ and $(B, C)$ have signed agreements, but the dyad $(A, B)$ continues fighting. This observed network of `peaceful' ties could be generated from any of  three unobserved sets of relations: it may be that $B$ failed to reciprocate $A$'s pursuit of peace, or vice versa, or neither $A$ nor $B$ pursued peace. To identify conditions that drive factions to sign peace agreements we must account for these unobserved asymmetries.  Here, the observed symmetric graph is an incomplete representation of the underlying, asymmetric network, which is frequently the object of scientific interest.  We refer to this situation as ``partial observability.'' 

We present the partial observability generalized bilinear mixed effects model (P-GBME) to address this challenge.  The model is a synthesis of the generalized bilinear mixed effects (GBME) model \citep{hoff:2005} and the bivariate or ``partial observability'' probit model \citep{poirier:1980, przeworski:vreel:2002}. The model can probabilistically reconstruct the directed network from which the observed, undirected graph emerged. The model enables the study of network ties in a regression framework by accounting for interdependencies as well as unobserved asymmetries in network relations. The stochastic actor-oriented model (SAOM) for networks \citep{snijders:pickup:2017} also allows for partial observability. However, SAOM was designed to assess how specific network features (e.g., $k$-star triangles) give rise to an observed network. The latent network approach is not used to study the role of specific network statistics.  Rather, latent network models aim to account for broad patterns of network interdependence using a variance decomposition regression framework.\footnote{See \citet{minhas:etal:2016:arxiv} for detailed discussion. Other approaches based on generalized spatio-temporal dependence can also recover directed predicted probabilities \citet{franzese:etal:2012}.}

We illustrate the P-GBME model by applying it to the bilateral investment treaties (BIT) network for each year in 1990-2012. The model substantially improves predictive accuracy relative to both conventional logit and standard GBME. As important, P-GBME extracts new information about the factors that drive treaty preferences, identifies important structural changes in the network, and highlights possible ``hidden'' agreements that are easily overlooked when latent network asymmetries are ignored.

\section{The Model}

Building on the random utility framework \citep{mcfadden:1980}, we model an actor $i = 1, \ldots,  N$ as having net utility from forming a tie with another $j$: $z_{ij} = \mu_{ij} + \epsilon_{ij}$ with $\mu_{ij}$ representing the systematic component (that depends on observables) and $\epsilon_{ij}$ representing the stochastic error. To account for interdependencies in actors' utilities from having ties, we use the ``latent space'' approach \citep{hoff:2005} and model these utilities as follows:
\begin{align}    
\left ( \begin{array}{c}
         z_{ij} \\
         z_{ji} \end{array} \right ) 
         & \sim \mathcal{N}
        \left ( \begin{array}{c}
         \mu_{ij} + a_i + b_j +    \bm{u}'_i\bm{v}_j  \\
         \mu_{ji} + a_j + b_i +  \bm{u}'_j\bm{v}_i \end{array}, \left [ \begin{array}{cc}
        \sigma^2 & \rho \sigma^2 \\
         \rho \sigma^2 & \sigma^2 \end{array} \right] \right ). \label{eq:modz}
\end{align}
The correlation, $\rho$, captures the ``reciprocity'' between the utilities that actors derive from tie formation. Parameters $a_i$ and $b_j$ are sender- and receiver-specific random effects, respectively, and they capture second-order network dependencies. The vectors $\bm{u}_i$ and $\bm{v}_i$ represent the location of actor $i$ in the latent space of `senders' and `receivers,' respectively. These random effects capture higher-order dependencies in network ties: $i$ derives a large utility from forming a tie with $j$, if $i$'s location in the latent space of `senders' $\bm{u}_i$ is close to $j$'s location in the latent space of `receivers' $\bm{u}_j$ (so that the cross-product $\bm{u}'_i\bm{v}_j$ is large).\footnote{As in previous literature, these random effects are modeled as $(a_i, b_i) \sim \mathcal{N}(0, \bm{\Sigma_{ab}})$, $\bm{u}_i \sim \mathcal{N}_K(\bm{0}, \sigma_u^2\bm{I})$, and $\bm{v}_i \sim \mathcal{N}_K(\bm{0}, \sigma_v^2\bm{I})$, where $\bm{\Sigma_{ab}}$, $\sigma_u^2$, and $\sigma_v^2$ are unknown parameters. The choice of $K$, dimensionality of the latent space, is discussed supplementary materials.} We express the systematic components of actors' utilities as linear functions of predictors:
\begin{align}
    \mu_{ij} & = \bm{\beta}^{(s)}\bm{x}_i^{(s)} + \bm{\beta}^{(r)}\bm{x}_j^{(r)} + \bm{\beta}^{(d)}\bm{x}_{ij}^{(d)}, \label{eq:itoj}\\
    \mu_{ji} & = \bm{\beta}^{(s)}\bm{x}_j^{(s)} + \bm{\beta}^{(r)}\bm{x}_i^{(r)} + \bm{\beta}^{(d)}\bm{x}_{ji}^{(d)} \label{eq:jtoi}.
\end{align}

A researcher cannot directly observe net utility (the $z$'s).  We only observe agents' behaviors, in this case undirected \emph{bilateral} ties. A directional tie $i \to j$ is formed if and only if $i$'s net gain from doing so is strictly positive, $z_{ij} > 0$.  Accordingly, the bilateral tie $i \leftrightarrow j$ is formed if and only if \emph{both} actors derive a net positive payoff from having a tie so that $z_{ij} > 0$ and $z_{ji} > 0$. A researcher observes an undirected (bilateral) tie $y_{ij} = y_{ji} = \{0, 1\}$ arising from the following data generating process:
\begin{align}
y_{ij} = y_{ji} & =   \left\{ \begin{array}{ll}
         1 & \mbox{if $z_{ij} > 0$ and $z_{ji} > 0$},\\
         0 & \mbox{else}.\end{array} \right. \label{mod:y}
\end{align}

Under a standard identifying restriction $\sigma^2 = 1$, the model is a partially observable probit regression \citep{poirier:1980}, augmented with random-effects to capture unobserved heterogeneity and inter-dyadic dependencies. Vectors $\bm{x}_i^{(s)}$ and $\bm{x}_i^{(r)}$ represent the sender-specific and receiver-specific covariates, respectively. A model for the directional link $i \to j$ (eq. \ref{eq:itoj}), uses variables $\bm{x}_i$ as sender-specific predictors, but these same predictors become receiver-specific in the model for the directional link $j \to i$ (eq. \ref{eq:jtoi}). Vector $\bm{x}_{ij}^{(d)}$ contains dyad-specific variables. These dyad-specific variables might be symmetric, $x_{ij} = x_{ji}$ (e.g., distance between countries), or not (e.g., export-import).

A partially observed probit model requires at least one of the following identifying restrictions: (1) regression equations \ref{eq:itoj} and \ref{eq:jtoi} must have the same parameters and/or (2) one equation contains a predictor not included in another equation \citep{poirier:1980}. If the dyadic predictors are asymmetric, $\bm{x}_{ij} \neq \bm{x}_{ji}$, then condition (2) is satisfied. Furthermore, regression equations \ref{eq:itoj} and \ref{eq:jtoi} have the same parameters, and so condition (1) holds as well; thus, the above model is parametrically identified. However, we impose an additional restriction that $\rho = 0$. While this restriction is not required for parametric identification, \citet{rajbhandari:2014} showed that finite sample estimates of $\rho$ are sensitive to the starting values and generally cannot be treated as reliable.\footnote{The estimation algorithm provided with this paper allows $\rho$ to be estimated, but caution should be used when utilizing this option.}
 
We estimate the model in a Bayesian framework using Markov chain Monte Carlo.  In the supplemental materials give a more detailed exposition of the model, prior assumptions, the sampling algorithm. We also provide results from a simulation study demonstrating that the model successfully recovers known parameter values.

\section{Application: Bilateral Investment Treaties}

We apply the P-GBME model to the network of bilateral investment treaties (BITs) from 1990 to 2012 using the standard United Nations BIT database. There is a vibrant debate on whether BITs boost FDI \citep{jandhyala:etal:2011, simmons:2014,minhas:2016}, but a proper resolution of this debate requires a convincing empirical model of treaty formation \citep{rosendorff:shin:2012}. Partial observability is one key challenge in building such model: the observed network of signed bilateral treaties is symmetric, while the underlying preferences for these treaties are  asymmetric. 

We fit the P-GBME model separately for each year of data using a suite of covariates that closely follows  the existing empirical literature (see supplementary materials). Our model improves on the previous literature by accounting for both network interdependence and partial observability. 

\subsection{Predictive Performance}

In Table~\ref{tab:perfAssess} we compare the in-sample and out-of-sample predictive performance of the P-GBME to that of pooled probit, which assumes dyadic independence and ignores partial observability, and GBME, which models dyadic interdependencies, but not partial observability. The predictive accuracy of the GBME model in this case is similar to the pooled probit. Adding the partial observability component to the GBME model, however, produces an additional substantial improvement in the predictive accuracy as shown by all metrics for the P-GBME model.

\begin{table}[h!]
\centering
\begin{tabular}{lcccccc}
\toprule
& & \multicolumn{2}{c}{In-sample} & & \multicolumn{2}{c}{Out-of-sample} \\
\cmidrule{3-4}\cmidrule{6-7}
& & ROC & PR & & ROC & PR\\
\midrule
Pooled probit& & 0.75 & 0.48 &  & 0.73 & 0.44\\
GBME && 0.76 & 0.47 & & 0.77 & 0.48 \\
P-GBME && 0.90 & 0.71 & & 0.91 & 0.78 \\
\bottomrule
\end{tabular}
\caption{Predictive performance in BIT data: area under the receiver operating characteristic curve (ROC) and area under the precision recall curve (PR).}
\label{tab:perfAssess}
\end{table} 

\subsection{Regression Parameters}

Existing models, including GBME, estimate a single coefficient for each predictor. This assumes away the possibility that the same factor differentially ``affects'' $i$'s demand for a treaty with $j$ and $i$'s attractiveness to $j$.  The P-GBME recovers directed sender- and receiver-effects for node-level covariates. For instance, our estimates suggest that, countries faster growth in GDP per capita were no more inclined to sign BITs with others (sender effect).  But high-growth countries were more attractive BIT partners to others (receiver effect). Supplementary materials describes regression parameter estimates in detail.

\subsection{The Structure of Latent Treaty Preferences}

\begin{figure}[ht]
  \centering
  \begin{tabular}{cc}
    1995 & 2010 \\
    \includegraphics[width=.4\textwidth]{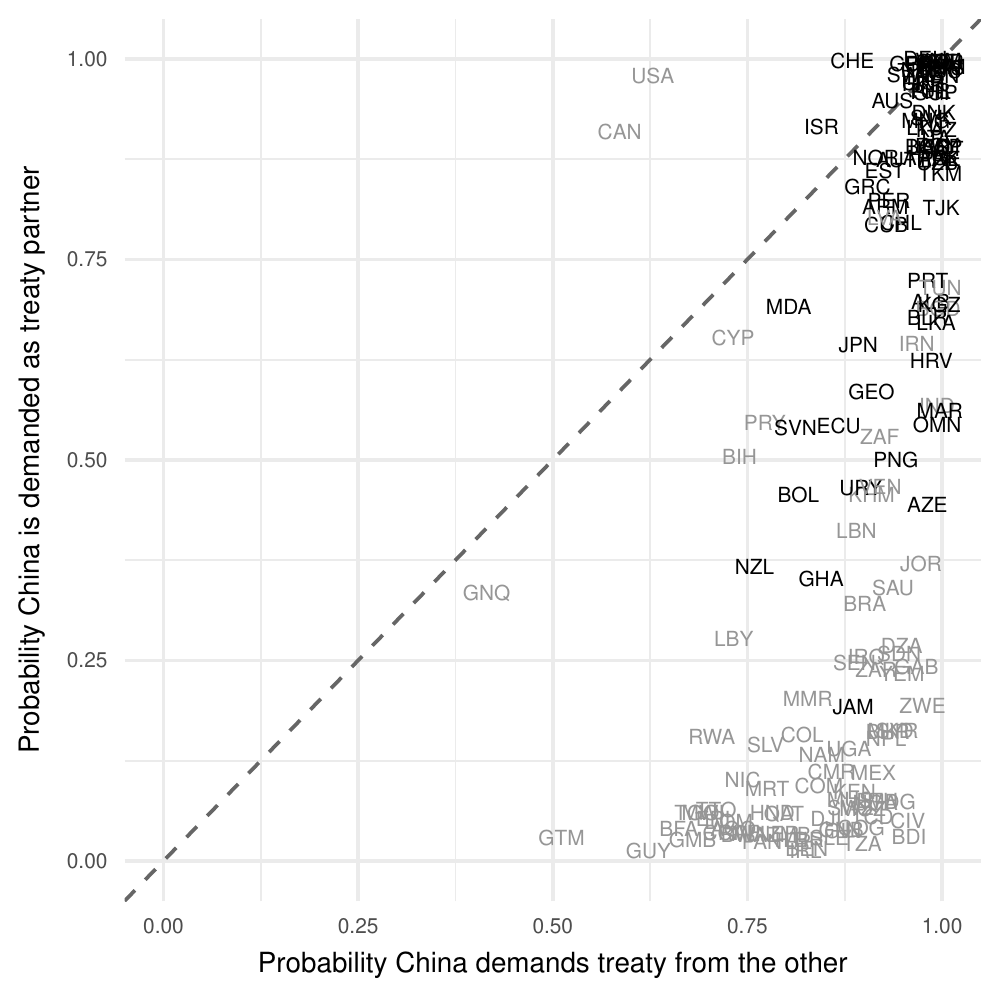} & 
    \includegraphics[width=.4\textwidth]{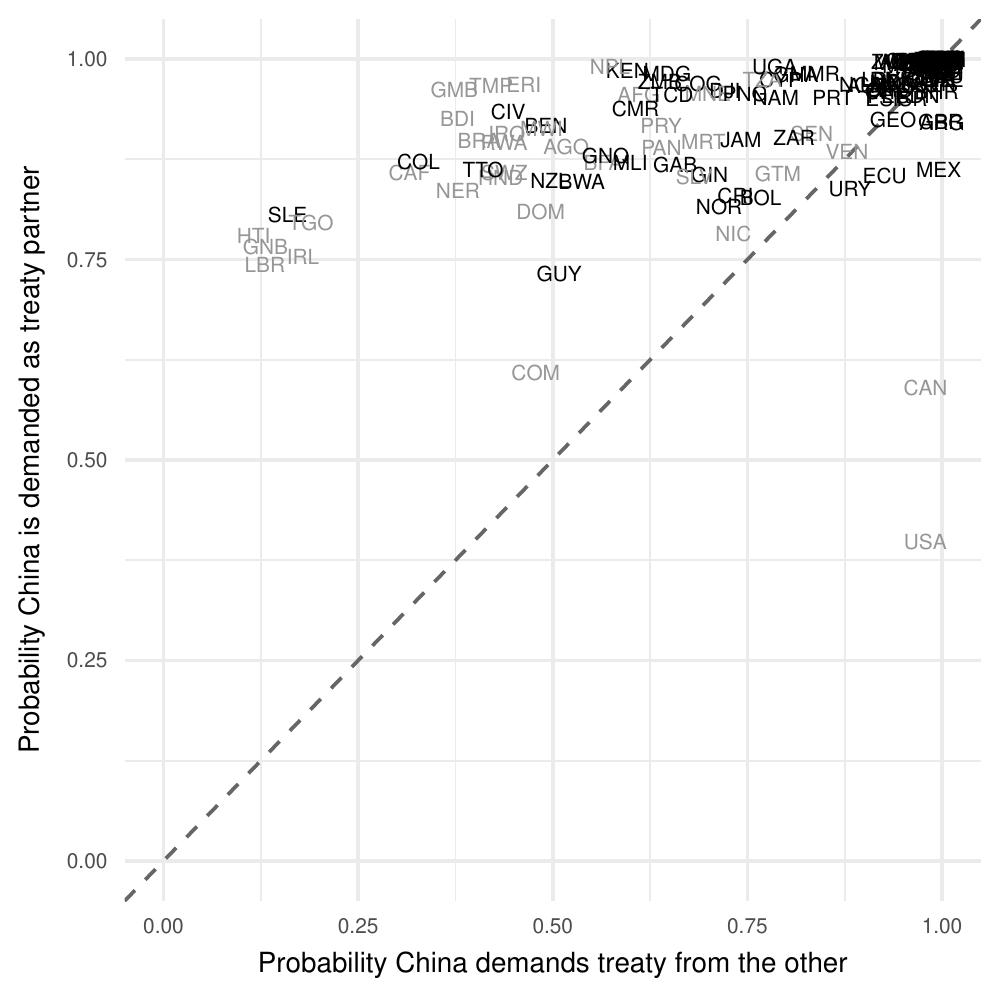} \\
  \end{tabular}
  \caption{Each panels displays the posterior mean probability that China demands and will be demanded as a BIT partner. Country labels in black designate those that had formed a BIT with China by the specified year.} 
  \label{fig:predplotCHN}
\end{figure}

The P-GBME model allows us to extract ``latent preferences'' for treaty formation -- the estimated probability that country $i$ demands a treaty from $j$, and vice versa. Figure~\ref{fig:predplotCHN} displays the mean posterior predicted probabilities relevant to China in 1995 and 2010. The horizontal axis represents a country's attractiveness to China as a BIT partner and the vertical axis is China's attractiveness to that country. Countries above the diagonal line find China a more attractive BIT partner than China finds them, and vice versa.  Color identifies observed BITs.

The plots reveal how China's position in the BIT network changed over time. In 1995, China was moving aggressively to demand BITs around the world, forming ties that the model views as relatively unlikely. By 2010, China had many more BITs in place and is more likely to be a treaty target by the remaining countries, an indication of China's expanded role in the global economy.  

\FloatBarrier

\begin{figure}[ht]
  \centering
\includegraphics[width=.7\textwidth]{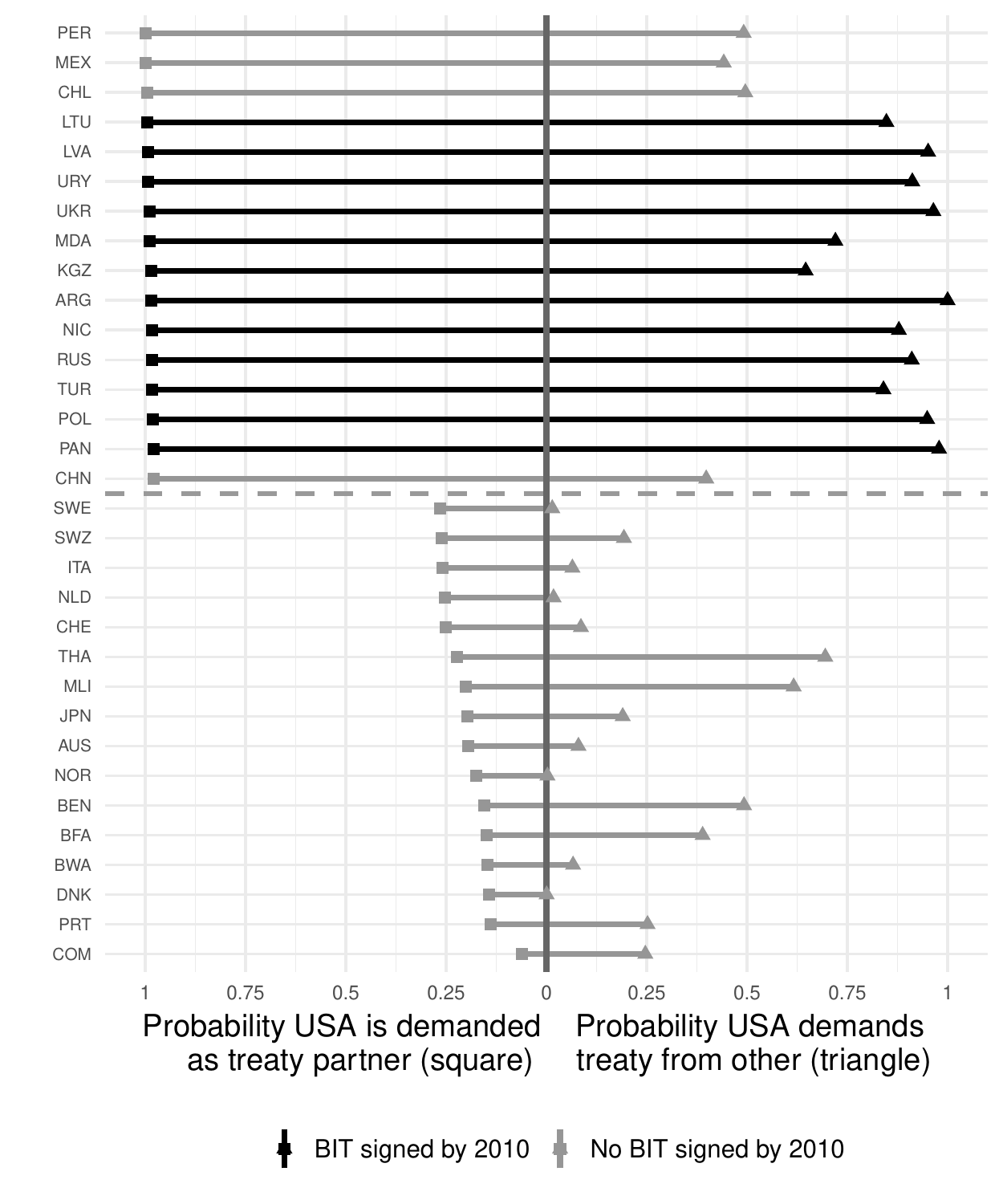} 
\caption{The posterior probability a country demands a BIT (left) and is demanded (right) with the USA in 2010 (the top and bottom decile of countries).  P-GBME is good at separating likely from unlikely treaties as well as identifying ``hidden agreements.''} 
  \label{fig:predplotUSA}
\end{figure}

Figure~\ref{fig:predplotUSA} illustrates a different use of the P-GBME model.  It shows the predicted probabilities that the USA is demanded (left) and demands (right) a BIT in 2010.  Observe that the model predicts Peru, Mexico, and Chile---the top 3 non-BITs---demand a BIT with the USA at probabilities close to one, and are also likely BIT targets of the USA with probabilities exceeding $0.5$. Closer inspection of UN treaty data reveals that, by 2010, all three had signed other agreements with the US that contain provisions functionally equivalent to BITs; these agreements do not appear in the BIT dataset commonly used in the literature.\footnote{The treaties are 2006 Peru-USA Free Trade Agreement (FTA), NAFTA in 1992, and the 2003 Chile-USA FTA. These agreements included investment provisions that mirror the terms of a BIT almost exactly (see \url{http://investmentpolicyhub.unctad.org/Download/TreatyFile/5454}).} The P-GBME model nevertheless highlights these ``hidden'' agreements as dyads likely to have a BIT. This suggests that researchers studying BITs need to carefully examine the dataset they employ and perhaps expand the set of treaties considered relevant.

\FloatBarrier

\section{Conclusion}

Partial observability occurs whenever the observed graph is undirected yet the underlying process implies directed relationships. We introduced a model that can reconstruct the latent directed network ties, and illustrated its advantages on an example of bilateral trade agreements. The future work in this area could focus on several extensions. First, we accounted for network dependencies using bilinear mixed effects framework (GBME), which could be generalized using  the recently developed additive and multiplicative effects network model \citep{hoff:2015:arxiv,minhas:etal:2016:arxiv}. Second, estimating this type of network model in a fully dynamic setting (as opposed to slicing data by time, as we did here) remains a challenge, especially when the set of nodes changes in time.

\clearpage

\appendix
\renewcommand{\thefigure}{\arabic{figure}}
\setcounter{figure}{0}
\renewcommand{\thetable}{\thesection \arabic{table}}
\setcounter{table}{0}
\setcounter{section}{0}
\appendix

\section*{\textbf{Supplemental Materials}}

\section{Details on the P-GBME}

As detailed in the paper, the partial observability generalized bilinear mixed effects (P-GBME) framework treats the observed symmetric outcome, $y_{ij} = y_{ji}$, as resulting from a joint decision taken by a pair of actors. We formalize the joint decision making process using a bivariate probit model with a standard normal link function:

\begin{align}
y_{ij} = y_{ji} & =   \left\{ \begin{array}{ll}
         1 & \mbox{if $z_{ij} > 0$ and $z_{ji} > 0$},\\
         0 & \mbox{else},\end{array} \right. \label{mod:y}\\
    \left ( \begin{array}{c}
         z_{ij} \\
         z_{ji} \end{array} \right ) 
         & \sim \mathcal{N}
        \left ( \begin{array}{c}
         \mu_{ij} + a_i + b_j +    \bm{u}'_i\bm{v}_j  \\
         \mu_{ji} + a_j + b_i +  \bm{u}'_j\bm{v}_i \end{array}, \left [ \begin{array}{cc}
        \sigma^2 & \rho \sigma^2 \\
         \rho \sigma^2 & \sigma^2 \end{array} \right] \right ), \label{eq:modz}\\
        (a_i, b_i)' & \sim \mathcal{N}\left(\bm{0}, \bm{\Sigma}_{ab} \right),\\
         \bm{u}_i & \sim \mathcal{N}_K(\bm{0}, \sigma_u^2\bm{I}),\\
         \bm{v}_i & \sim \mathcal{N}_K(\bm{0}, \sigma_v^2\bm{I})\label{eq:modv}.
\end{align}

$a_{i}$ and $b_{j}$ represent sender and receiver random effects that account for first order dependence patterns that often arise in relational data, while $\bm{u}'_i\bm{v}_j$ captures the likelihood of a pair of actors interacting with one another based on third order dependence patterns such as transitivity, balance, and clustering. For identification purposes, we fix $\sigma^2=1$ and $\rho=0$. The former is a standard restriction in probit frameworks with a binary outcome. We undertake the latter restriction because \citet{rajbhandari:2014} shows that in this framework it is difficult to recover reliable estimates for $\rho$ as the parameter is highly sensitive to the initial value. 

The sender and receiver random effects ($a_{i}$ and $b_{j}$) are drawn from a multivariate normal distribution centered at zero with a covariance matrix, $\Sigma_{ab}$, parameterized as follows:

\begin{align}
  \Sigma_{ab} = \begin{pmatrix} \sigma_{a}^{2} & \sigma_{ab} \\ \sigma_{ab} & \sigma_{b}^2   \end{pmatrix}
\end{align}

The nodal effects are modeled in this way to account for the fact that in many relational datasets we often find that actors who send a lot of ties are also more likely to receive a lot of ties. Heterogeneity in the the sender and receiver effects is captured by $\sigma_{a}^{2}$ and $\sigma_{b}^{2}$, respectively, and $\sigma_{ab}$ describes the covariance between these two effects. 

$\mu$ represents the systematic component of actors' utilities and is expressed as a linear function of sender $^{(s)}$, receiver $^{(r)}$, and dyadic $^{(d)}$ covariates: 

\begin{align}
    \mu_{ij} & = \bm{\beta}^{(s)}\bm{x}_i^{(s)} + \bm{\beta}^{(r)}\bm{x}_j^{(r)} + \bm{\beta}^{(d)}\bm{x}_{ij}^{(d)}, \label{eq:itoj}\\
    \mu_{ji} & = \bm{\beta}^{(s)}\bm{x}_j^{(s)} + \bm{\beta}^{(r)}\bm{x}_i^{(r)} + \bm{\beta}^{(d)}\bm{x}_{ji}^{(d)} \label{eq:jtoi}.
\end{align}

This formulation allows us to incorporate exogenous actor and dyad level characteristics into how actors make decisions within the partial probit framework. Following \citet{hoff:2005}, to enable a more efficient estimation, we reparameterize the model to implement hierarchical centering of the random effects \citep{gelfand:etal:1995}:

\begin{align}
  z_{i,j} & \approx \mathcal{N}(\bm{\beta}^{(d)}\bm{x}_{ij}^{(d)} + s_{i} + r_{j} + \bm{u}'_i\bm{v}_j), \\
  s_{i} & = \bm{\beta}^{(s)}\bm{x}_i^{(s)} + a_{i}, \\
  r_{j} & = \bm{\beta}^{(s)}\bm{x}_j^{(r)} + b_{j}.
\end{align}

\subsection{Parameters and Priors}


To estimate the parameters discussed in the previous section, we utilize conjugate priors and a Monte Carlo Markov Chain (MCMC) algorithm. Prior distributions for the parameters are specified as follows:\footnote{For details on the full conditional distributions of each of the parameters see \citet{hoff:2005}.} 

\begin{itemize}
  \item $\bm{\beta}^{(s)}$, $\bm{\beta}^{(r)}$, and $\bm{\beta}^{(d)}$ are each drawn from multivariate normals with mean zero and a covariance matrix in which the covariances are set to zero and variances to 10
  \item $\Sigma_{a,b} \sim \text{ inverse Wishart}(I_{2\times 2}, 4)$
  \item $\sigma_u^2$, and $\sigma_v^2$ are each drawn from an i.i.d. inverse gamma(1,1).
\end{itemize}

Starting values for each of the parameters are determined using maximum likelihood estimation.

\subsection{The MCMC algorithm}

To estimate this model a Gibbs sampler is used. This sampler follows the procedure laid out in \citet{hoff:2005,hoff:2009} with the exception of the first step in which we extend the GBME by accounting for the possibility that seemingly symmetric events are the result of a joint decision between a pair of actors. This first step involves sampling from a truncated normal distribution, we show the full conditional distribution below.

\begin{enumerate}
  \item Modeling partially observable outcome. Conditional on there being an observed link between $i$ and $j$, and conditional on other parameters, we draw the latent variables $z_{ij}$ and $z_{ji}$ from the bivariate normal distribution such that both latent variables are positive:
\begin{align*}
\left(
\begin{matrix}
  z_{ij} \hfill \\   z_{ji}
\end{matrix}
\, \middle\vert \,
y_{ij} = 1
\right) & \sim \mathcal{N}
        \left ( \begin{array}{c}
         \mu_{ij} + a_i + b_j +    \bm{u}'_i\bm{v}_j  \\
         \mu_{ji} + a_j + b_i +  \bm{u}'_j\bm{v}_i \end{array}, \left [ \begin{array}{cc}
        \sigma^2 & \rho \sigma^2 \\
         \rho \sigma^2 & \sigma^2 \end{array} \right] \right )\I\{z_{ij} > 0 \cap z_{ji} > 0\}.
\end{align*}
Conditional $y_{ij} = 0$ (there is no observed link between $i$ and $j$), we sample the latent variables from the bivariate normal distribution where at least one of the latent variables, $z_{ij}$ or $z_{ji}$, is constrained to be negative:
\begin{align*}
\left(
\begin{matrix}
  z_{ij} \hfill \\   z_{ji}
\end{matrix}
\, \middle\vert \,
y_{ij} = 0
\right) & \sim \mathcal{N}
        \left ( \begin{array}{c}
         \mu_{ij} + a_i + b_j +    \bm{u}'_i\bm{v}_j  \\
         \mu_{ji} + a_j + b_i +  \bm{u}'_j\bm{v}_i \end{array}, \left [ \begin{array}{cc}
        \sigma^2 & \rho \sigma^2 \\
         \rho \sigma^2 & \sigma^2 \end{array} \right] \right )\I\{z_{ij} < 0 \cup z_{ji} < 0\}.
\end{align*}
  \item Additive effects
  \begin{itemize}
    \item Sample $\bm{\beta}^{(d)}, s, r \;|\; \bm{\beta}^{(s)}, \bm{\beta}^{(r)}, \Sigma_{a,b}, Z, \bm{U}, \bm{V}$ (linear regression)
    \item Sample $\bm{\beta}^{(s)}, \bm{\beta}^{(r)} \;|\; s, r, \Sigma_{a,b}$ (linear regression)
    \item Sample $\Sigma_{a,b}$ from full conditional distribution
  \end{itemize}
  \item Multiplicative effects\footnote{See \citet{hoff:2009} for further details on how multiplicative effects are estimated in a directed context within the GBME framework. }
  \begin{itemize}
    \item For $i = 1, \ldots, n$:
    \begin{itemize}
      \item Sample $u_{i} \;|\; \{u_{j}, j \neq i\}, Z, \bm{\beta}^{(d)}, s, r, \sigma_u^2, \sigma_v^2, \bm{V}$ (linear regression)
      \item Sample $v_{i} \;|\; \{v_{j}, j \neq i\}, Z, \bm{\beta}^{(d)}, s, r, \sigma_u^2, \sigma_v^2, \bm{U}$ (linear regression)
    \end{itemize}
  \end{itemize}
\end{enumerate}

\subsection{Simulation Exercise}

To test the capabilities of the P-GBME framework in representing the data generating process for a partially observable outcome we conduct a simulation exercise. In each simulation, we randomly construct a directed network from a pair of dyadic covariates, nodal covariates, and the random effects structure detailed in the previous section. The regression parameters for the dyadic covariates are set at 1 and -1/2, and the parameters for the nodal covariates are set at 0 and 1/2. At this stage, the network simulated from this data generating process is directed. We modify the simulated network so that a link between a dyad only appears in the network if both the $i,j$ dyad and the $j,i$ dyad both have a link in the simulated network, thus making the network appear undirected. 

Next, we examine whether the P-GBME model can recover the data generating process underlying the partially observed simulated network. We run the P-GBME model in every simulation for 20,000 iterations with a 10,000 burn-in period. We repeat this simulation process 100 times.

\begin{figure}[ht]
  \centering
  \includegraphics[width=1\textwidth]{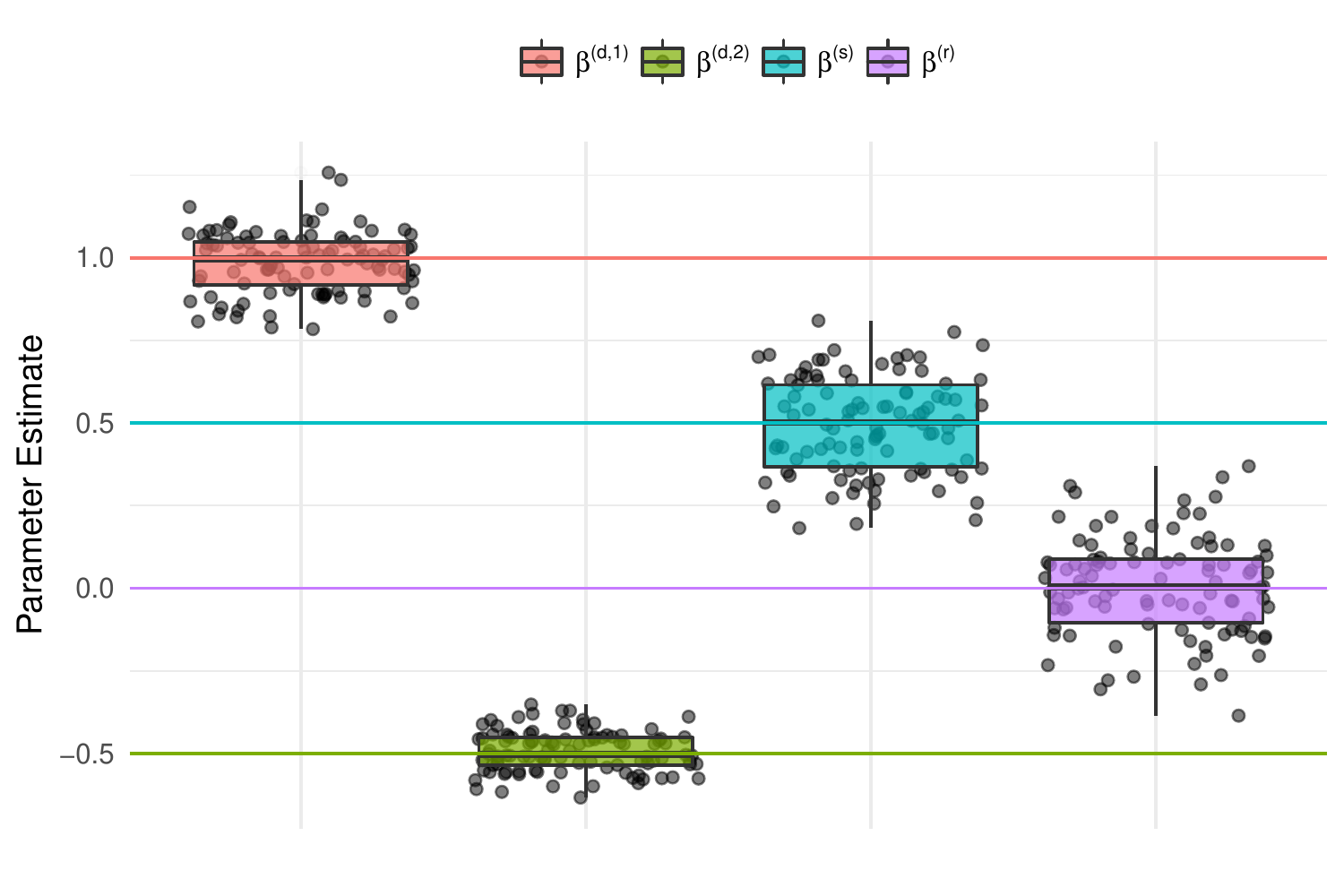}
  \caption{Boxplot of mean value of regression parameters estimated using the P-GBME across 100 simulations. Horizontal colored lines indicate the true parameter values.}
  \label{fig:simBias}
\end{figure}

With the simulation results our first step is to examine whether the P-GBME accurately recovers the regression parameter estimates for the dyadic and nodal covariates. To test whether this is the case we calculate the mean regression parameter estimate from the MCMC results for each simulation, and we summarize these results in Figure~\ref{fig:simBias}. For each parameter we indicate its true value by a colored horizontal line and summarize the distribution of the mean regression values estimated from the P-GBME using a boxplot. Given that for each of the parameters the true value almost exactly crosses the median value indicated in the box plot, this simulation shows that the P-GBME is quite effective in estimating the true parameter values underlying a partially observed outcome. 

\FloatBarrier

We also examine the proportion of times that the true value falls within the 95\% credible interval of the estimated regression parameter. In over ninety percent of the simulations, the true value falls within the 95\% credible interval of each of the estimated regression parameters from the P-GBME. Specifically, for $\beta^{(d,1)}$ the coverage rate is 0.85, for $\beta^{(d,2)}$ 0.93, for $\beta^{(s)}$ 0.93, and for $\beta^{(r)}$ 0.97.

\section{Estimation and Application}

\subsection{Data}

Table~\ref{tab:vars} provides a description for each of the variables used in the analysis.

\begin{table}[ht]
\footnotesize
  \centering
  \caption{Variables in the analysis}\label{tab:modelSpec}
  \begin{tabular}{llp{7cm}p{4cm}}  
    \toprule
    Variable & Level & Definition & Source \\
    \midrule
    BIT & dyadic & $1$ if $i$ \& $j$ Signed a BIT by year $t$ & UNCTAD\footnote{\url{http://investmentpolicyhub.unctad.org/IIA}}\\
    UDS (median) & dyadic & $|$UDS$_{it} - $UDS$_{jt}|$ &Pemstein et al. (2010)\nocite{pemstein:etal:2010}  \\
    Law \& Order & dyadic  &$|$LO$_{it} - $LO$_{jt}|$& ICRG\footnote{\url{http://epub.prsgroup.com/products/international-country-risk-guide-icrg}}\\
    Log(GDP per capita)  &dyadic& $|$Log(GDPcap)$_{it} - $Log(GDPcap)$_{jt}|$ & WDI \\
    OECD & dyadic& $1$ if $i$\& $j$ Both OECD members by year $t$&OECD \\
    Distance &dyadic&Minimum distance between $i$\&$j$ & Gleditsch \& Ward (2001) \nocite{gleditsch:ward:2001}\footnote{Also see \citet{weidmann:gleditsch:2015}.}\\
    FDI/GDP &node&Net FDI inflow as \% GDP in year $t$&WDI\\
    ICSID Disputes &node& Cumulative number of disputes by year $t$& ICSID\footnote{\url{https://icsid.worldbank.org/en/Pages/cases/AdvancedSearch.aspx}} \\
    GDP per capita growth&node& Level of GDP per capita growth by year $t$ &WDI\\ 
    PTAs&node& Cumulative number of PTAs signed by year $t$& DESTA\footnote{\url{https://www.designoftradeagreements.org/}}\\
    \bottomrule
  \end{tabular}
  \label{tab:vars}
\end{table}
\FloatBarrier

A shortcoming of the existing GBME framework is its inability to account for applications where there is missingness in the set of exogenous covariates used in the model. For our application, a number of the nodal covariates had varying levels of missingness. Additionally, most of the dyadic covariates that we construct from nodal variables, such as the unified democracy scores, also have varying levels of missingness. The table below shows how much missingness we had for the variables included in our analysis: 

\begin{table}[ht]
  \centering
  \caption{Missingness among variables used in the analysis}
  \begin{tabular}{lc}
  \toprule
  Variable & Proportion of Cases Missing \\
    \midrule
    Law \& Order & 16.6\% \\ 
    FDI/GDP & 2.8\% \\
    GDP per capita & 1.4\% \\
    GDP per capita growth & 1.4\% \\
    Unified Democracy Scores (UDS) & 0.7\% \\
    ICSID Disputes & 0\% \\
    PTAs & 0\% \\   
    OECD & 0\% \\
    \bottomrule
  \end{tabular}
\end{table}
\FloatBarrier

In general, the level of missingness is not high. The only exception here is with the Law \& Order variable from the ICRG dataset, for this variable we had approximately 17\% of country-year observations missing from 1990 to 2012. The only true dyadic variable we include in our analysis is a calculation of the minimum distance between countries, and this variable has no missingness. Additionally, our dependent variable measuring whether or not two countries had signed a BIT by year $t$ also has no missingness. 

A number of works have noted the issues that can arise when simply using listwise deletion,\footnote{See, for example, \citet{king:etal:2001}.} thus before running the P-GBME sampler we impute missingness among the covariates used in our model with a Bayesian, semi-parametric copula imputation scheme.\footnote{See \citet{hoff:2007a,hollenbach:etal:2016a} for details on this imputation scheme and how it differs from other approaches frequently utilized in political science.} We generate 1,000 imputed datasets from this imputation scheme and save 10 for use in the P-GBME MCMC sampler.

To account for missingness within the P-GBME, at the beginning of every iteration of the MCMC for model, we draw a randomly sampled imputed dataset from the posterior of the Copula, calculate the parameters associated with the P-GBME using the imputed dataset, and repeat this process for every iteration of the sampler for the model. This approach directly incorporates imputation uncertainty into our posterior distributions of the P-GBME parameters without having to run and combine separate models. 

\subsection{Estimation details}

In our application we estimate the P-GBME separately for each year from 1990 to 2012 using the prior distributions and MCMC algorithm described above.  For each year, we ran the P-GBME MCMC sampler for 20,000 iterations, discarding the first 10,000 iterations as burn-in. We thinned the chain by saving only every 10$^{th}$ value. 

The following trace plot describes MCMC convergence for all parameters in the 2012 P-GBME model.  

\begin{figure}[ht]
\centering
\includegraphics[width=1\textwidth]{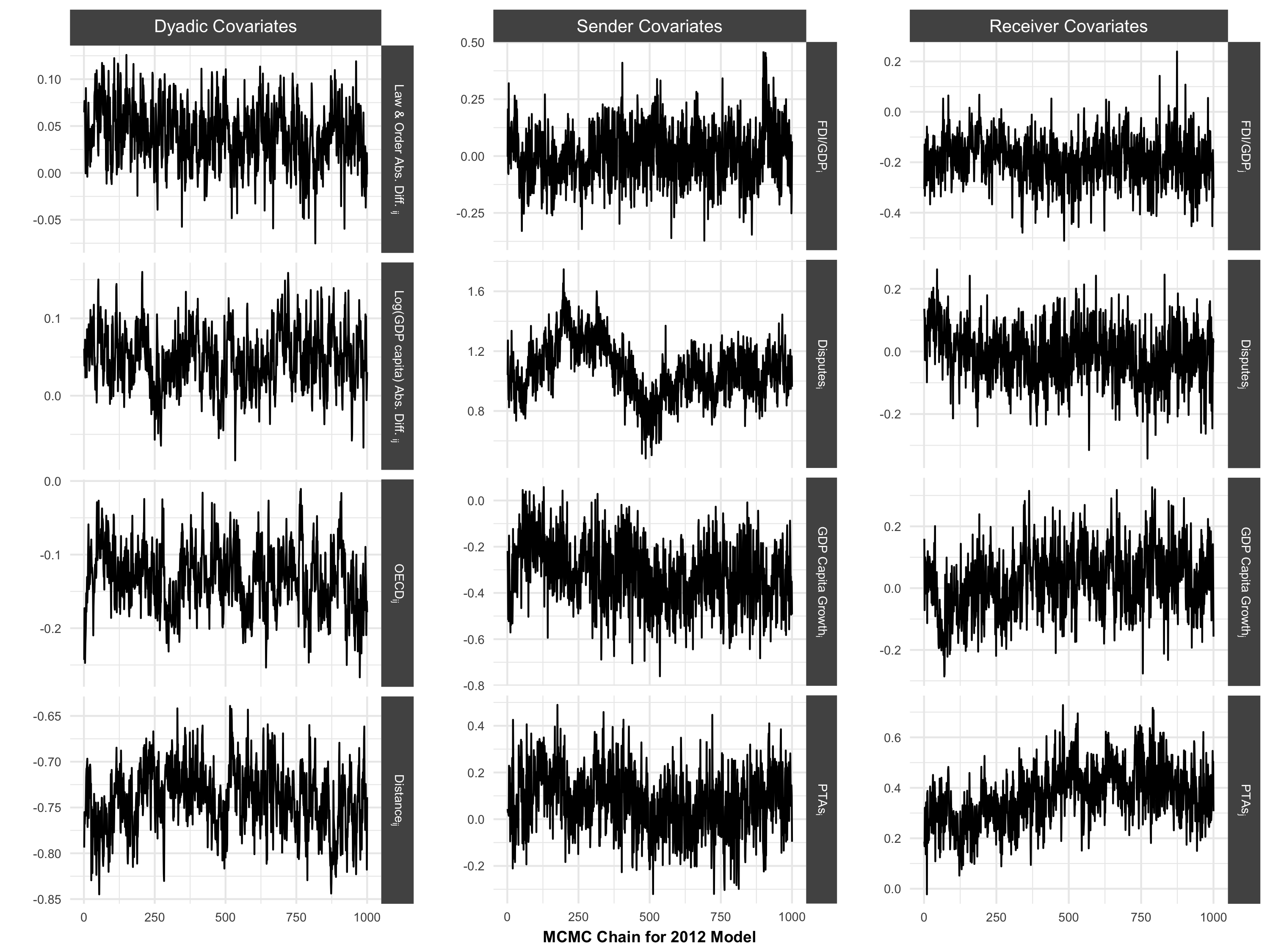}
\caption{Traceplot for 2012 P-GBME results.}
\label{fig:trace}
\end{figure}
\FloatBarrier

\subsection{Regression Parameters}

\begin{figure}[ht]
  \centering
  \includegraphics[width=1\textwidth]{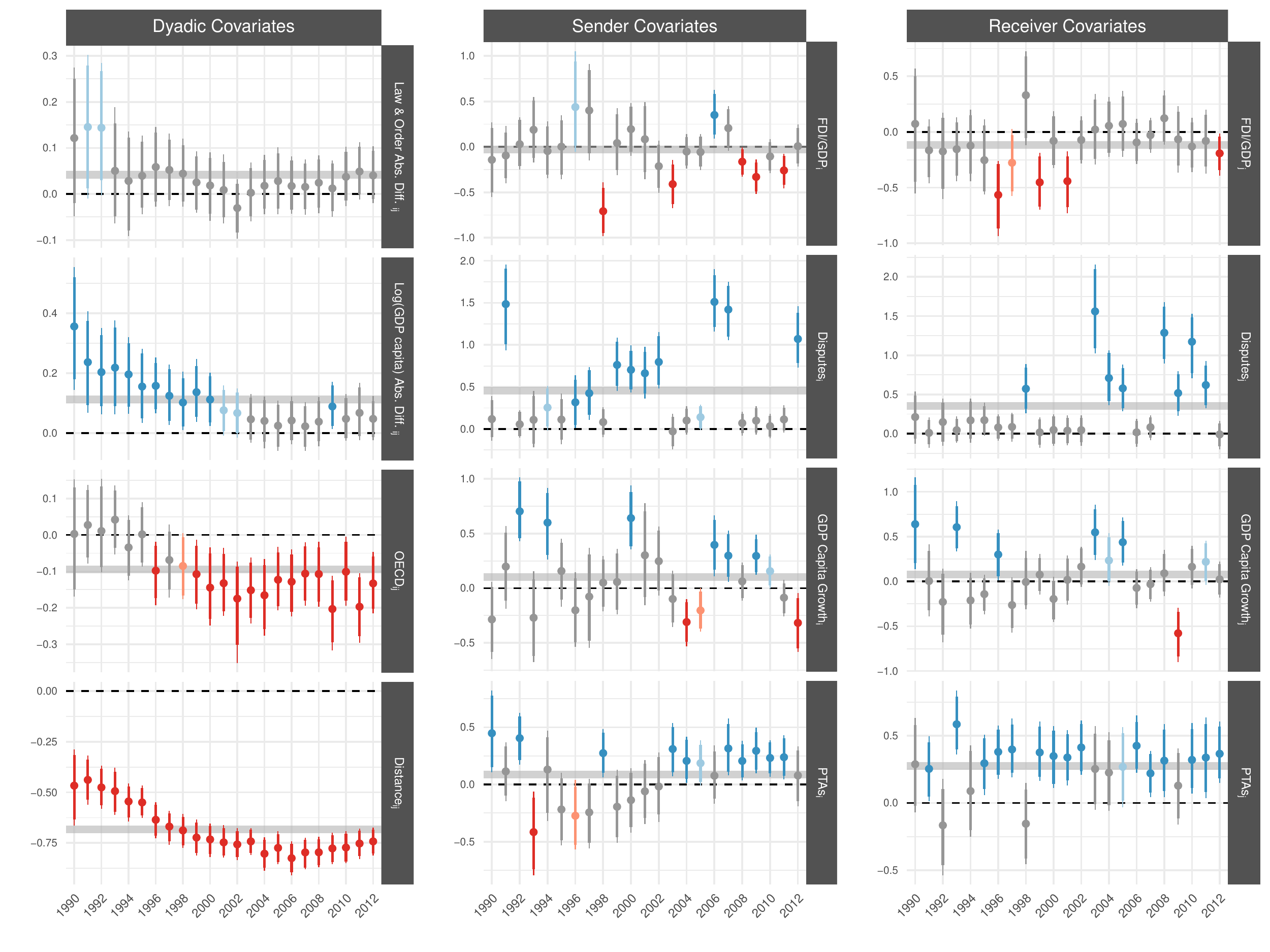}
  \caption{Left-most plot shows results by year for the dyadic parameters, next shows parameter results for sender covariates, and right-most plot show results for receiver covariates. Points in each of the plots represents the average effect for the parameter and the width the 90 and 95\% credible intervals. The grey bar in the panels represents the average effect of the parameter across all years. Dark shades of blue and red indicate that the 95\% CI does not contain 0 and lighter shades implies that the $0$ is not in the 90\% credible interval. Parameters in grey are are ones where 0 is inside the 90\% credible interval.} 
  \label{fig:coef}
\end{figure}

Figure~\ref{fig:coef} displays the posterior mean and the 90\% (thicker) and 95\% (thinner)  credible intervals (CIs). The first column contains the dyadic covariates; the second, sender-level covariates; and, the third, receiver covariates. In each panel, we show the parameter estimates for that variable from 1990 to 2012.  The dotted horizontal line is $0$ and the thicker grey line is the posterior mean, pooling the posterior draws across all years.   

The P-GBME recovers directed sender- and receiver-effects for node-level covariates from an observed undirected network, something that other approaches, including the GBME, are unable to do. Our estimation in this application indicates substantial instability in these estimates over time, both relative to a baseline of $0$ and relative to the pooled posterior mean.  This instability is consistent with substantive arguments that the incentives to sign BITs have changed over time \citep{jandhyala:etal:2011}. 

Dyadic covariates tend to be more stable across years in this application.  But they, too, show that the BIT formation process has changed over time.  Economic and political ``distance'', for example, has become less important as the network evolved and more lower-income countries have signed BITs with each other.  Geographic distance, on the other hand, continues to be strongly related to the formation of BITs.

P-GBME covariate estimates shed light on the evolving processes producing the observed BIT network in the 1990-2012 period.  The changing values of covariate parameters over time also indicates that common practice of pooling dyads and assuming the existence of temporally stable parameters may be dangerous.

\subsection{Choosing dimension of the multiplicative effects, $K$}

One of the parameters that users are able to set within the P-GBME to account for third order dependence patterns is $K$ -- see the MCMC algorithm section above for more details on this parameter and its relation to the model. In the results reported in the paper, we set $K$ = 2. To understand whether or not a higher value of $K$ is necessary users of this approach can compare the in-sample fit of the model with varying values for $K$. In our application exercise, we varied $K$ from 1 to 3 to settle on an appropriate value of $K$ that can represent the data generating process of the network. Results are shown in Table~\ref{tab:kvar} below.

\begin{table}[ht]
\centering
\caption{In-sample performance results from running the P-GBME on data from 2012 with varying values of $K$.}
\label{tab:kvar}
\begin{tabular}{lcc}
\toprule
~ & AUC (ROC) & AUC (PR) \\
\midrule
  $K$=1 & 0.87 & 0.63 \\
  $K$=2 & 0.90 & 0.71 \\
  $K$=3 & 0.92 & 0.71 \\ 
\bottomrule
\end{tabular}
\end{table}

As you can see after $K$=2, the subsequent in-sample performance improvement notably declines. There is a slight increase in performance from $K$=2 to $K$=3, however, every time one increases $K$ we are also adding $2*n$ more parameters to the P-GBME model. Adding this many more parameters can easily lead one to overfit the data in an out-of-sample context. 

\clearpage
\singlespacing
\bibliographystyle{apsr}
\bibliography{master}

\end{document}